\documentclass{emulateapj}
\usepackage{epsfig,graphicx}

\def\beq{\begin{eqnarray}}
\def\eeq{\end{eqnarray}}
\def\acf{w(\theta)}
\def\deg2{\, {\rm deg}^{-2}}
\def\gam{-0.79\pm0.02}
\def\A{0.10\pm0.01}

\begin{document}

\title{The Galaxy Angular Correlation Functions and Power Spectrum from 
the Two Micron All Sky Survey}

\author{Ariyeh H. Maller, Daniel H. McIntosh, Neal Katz and Martin D. Weinberg}
\affil{Astronomy Department, University of Massachusetts, Amherst, MA 01003}

\begin{abstract}
We calculate the angular correlation function of galaxies in the 
{\it Two Micron All Sky Survey}.  We minimize the possible contamination by
stars, dust, seeing and sky brightness by studying their cross correlation
with galaxy density, and limiting the galaxy sample accordingly.
We measure the correlation function at scales between 
$1\arcmin < \theta < 18\arcdeg$ using a half million galaxies. We find a best 
fit power law to the correlation function has a slope of $\gam$ and an 
amplitude at $1\arcdeg$ of $\A$ in the range $1\arcmin - 2.5\arcdeg$. However,
there are statistically significant oscillations around this power law.
The largest oscillation occurs at about 0.8 degrees,
corresponding to $700h^{-1}$ kpc at the median redshift of our survey,
as expected in halo occupation distribution descriptions of galaxy clustering.
In addition, there is a break in the power-law shape of the correlation 
function at $\theta > 2.5\arcdeg$. Our results are in good agreement with 
other measurements of the angular correlation function. 

We invert the angular correlation function using Singular Value Decomposition
to measure the three-dimensional power spectrum and find that it too
is in good agreement with previous measurements.  A dip seen in the power
spectrum at small wavenumber $k$ is statistically consistent with CDM-type 
power spectra. A fit of CDM-type power spectra
in the linear regime ($k < 0.15 h$ Mpc$^{-1}$)
give constraints of $\Omega_m h = 0.13\pm 0.07$ 
and $\sigma_8=1.0\pm0.09$ for a spectral index of $1.0$. 
This suggest a $K_s$-band linear bias of $1.1\pm0.2$.
These measurements are in good agreement with other measurements 
of the power spectrum on linear scales.
On small scales the power-law shape of our power spectrum is shallower 
than that derived for the {\it Sloan Digital Sky Survey}. This may imply a
biasing for these different galaxies that could be either waveband or 
luminosity dependent.  The power spectrum derived here in combination with 
the results from other surveys can be used to constrain models of galaxy 
formation. 
\end{abstract}


\section{Introduction}
Correlation statistics are an important method for relating galaxies
to the underlying mass distribution.  The angular correlation 
function, $\acf$, measures the projected clustering of galaxies by
comparing the distribution of galaxy pairs relative to that of 
a random distribution.
While a less direct probe than the three-dimensional 
correlation function $\xi(r)$, the angular correlation function  can 
be a powerful approach owing to the larger sizes of two-dimensional surveys.
The angular correlation function for bright galaxies has been measured 
for the Lick Survey \citep{gp:77}, the Automated Plate Measuring (APM) 
Galaxy Survey \citep{madd:90,mes:96}, the Edinburgh Durham Southern (EDS) 
Galaxy Catalog \citep{cnl:92}, and the Muenster Red Sky Survey 
\citep{bosch:02}. Most recently the Sloan Digital Sky Survey 
\citep[SDSS;][]{york:00} has performed a very detailed 
analysis of the angular correlation function \citep{conn:02,scra:02} 
from their Early Data Release \citep[EDR;][]{stou:02}. 

Additionally, numerous papers have
been written on techniques for inverting the angular correlation function
to determine the full three-dimensional power spectrum, $P(k)$ 
\citep{limber:53,lucy:74,be:93,dg:00,em:01,ez:01,pb:03}.

In this paper we calculate the angular correlation function for galaxies
selected from the {\it Two Micron All Sky Survey} \citep[2MASS;][]{skrut:97}. 
We then invert $\acf$ using Singular Value Decomposition, as suggested 
by \citet{ez:01}, to measure the three-dimensional power spectrum.  
A preliminary analysis of $\acf$ and $P(k)$ for the Second Incremental 
Data Release of the 2MASS catalog was performed by \citet{abp:01}.  
Here we use the complete and final 2MASS catalog, which provides a more 
than two-fold increase in the number of sources owing to the nearly full sky 
\citep[$>98\%$,][]{jarr:00a} coverage, and
removes the significant complication of coverage masking suffered by
\citet{abp:01}. In addition, the final 2MASS pipeline processing has
improvements in the flux measurements that affect the catalog completeness.  

The angular correlation
function of the 2MASS galaxies is interesting for a number of reasons.  First,
with 2MASS we can characterize the large-scale clustering of galaxies
in the near-infrared (NIR) using the $K_s$ (2.15\micron) passband.  
Galaxy $K_s$ light is 5-10 times less sensitive to dust and
stellar populations than $B$-band light, providing a more uniform survey
of the galaxy population.  Because
the $K_s$-band most closely measures stellar mass, it is possible that 
the $K_s$ selected correlation function is a better probe of the dark matter
power spectrum than $B$ and $R$-band selected measurements.
The variations of correlation functions in
different bands contain information about how galaxy properties are related
to the underlying dark matter distribution and, therefore, to their formation
and evolution.

Secondly, 2MASS has full sky coverage extending out to a median redshift of 
$z=0.07$ and, hence, measures the power spectrum of our local Universe.
Knowledge of our local ``cosmography'', and how it may differ from other 
regions of the Universe, is relevant for many cosmological tests.  
Thus the angular correlation function and power spectrum from 2MASS are 
valuable tools for understanding galaxy formation and cosmology especially
when used in comparison with measurements from other surveys. 

Our paper largely follows the treatment of the Sloan EDR performed by
\citet{scra:02}, \citet{conn:02} and \citet{dode:02} and is outlined as 
follows:  In \S\ref{sec:2mass} we describe our galaxy sample selection
from 2MASS.  In \S\ref{sec:cross} we analyze the importance of 
systematic errors by studying their cross correlation with galaxy density.  
In \S\ref{sec:acf} we calculate the correlation function and discuss 
its errors, and in \S\ref{sec:3D} we invert $\acf$ to measure the 
three-dimensional power spectrum, $P(k)$.  We conclude in \S\ref{sec:conc}. 

\section{2MASS Selected Galaxies}
\label{sec:2mass}
To measure the angular correlation function accurately, it is critically
important to fully understand the
reliability and completeness of the galaxy sample.
We select galaxies from the 2MASS extended source catalog 
\citep[XSC;][]{jarr:00a}, which contains over 1.1 million extended objects 
brighter than $K_s=14$ mag.\footnote{Completeness in the XSC is determined 
in terms of magnitudes 
measured inside the $20$~mag arcsec$^{-2}$ elliptical 
isophote (k\_m\_k20fe in the database).}
The detection of galaxies is limited predominantly by confusion
noise from stars, whose number density increases exponentially
towards $|b|=0\arcdeg$.  The XSC is mostly
galaxies at $|b|>20\arcdeg$ ($>98\%$), with
an increasing stellar mixture ($\sim10\%$) at lower
latitudes of $5\arcdeg < |b| < 20\arcdeg$. 
Within the Galactic plane, i.e. $|b|<5\arcdeg$,
there is an additional contamination by artifacts ($10-20\%$) and a variety of 
Milky Way extended sources ($\sim40\%$) including
globular and open clusters, planetary nebulae,
\ion{H}{2} regions, young stellar objects, nebulae, and giant molecular
clouds \citep{jarr:00b}.
For our final analysis, we employ a latitude cut
of $|b|>20\arcdeg$ to remove the strong contamination of $\acf$ from stars
at lower latitudes (see \S 3.1). Our cut reduces the Galactic
contaminant sources to $\sim2\%$ for our galaxy sample. 

The XSC also contains a fraction of spurious sources comprised 
of multiple star systems.  
Multiple star systems become an increasingly important source of contamination
as the stellar density, $n_{\rm star}$ rises,
where $n_{\rm star}$ is the number of $K_s<14$ mag stars per sq. degree 
calculated in a co-add which is $8.5\arcmin \times 16 \arcmin$.
Specifically, the reliability of separating stars from extended sources
drops rapidly from $95\%$ at $|b|=10\arcdeg$ 
($n_{\rm star}\approx 5,000 \deg2$), to $<65\%$ at $|b|<5\arcdeg$ 
($n_{\rm star}> 10^4 \deg2$) \citep{jarr:00b}.  
To remove sources that are certainly unreliable, we use a
cut at $n_{\rm star}< 10^4 \deg2$ during our initial galaxy selection 
from the XSC.
The stellar density is this high over $5.5\%$ of the sky so our galaxy catalog
only covers $94.5\%$ of the sky.
At high latitudes ($|b|>20\arcdeg$, 
$n_{\rm star}\lesssim 1,250 \deg2$) the catalog is
$>98\%$ reliable for sources with $K_s<13.5$ \citep{jarr:00a}.
The XSC contains a small fraction
of artifacts, {\it e.g.} diffraction spikes, meteor streaks,
infrared airglow, at all latitudes.  The XSC processing
removes and/or flags most artificial sources, as
described fully in \citet{jarr:00a}.  We use the XSC confusion flag   
(cc\_flag) to remove sources identified as artifacts.  We also remove
a small number of bright ($K_s<12$ mag) sources ($\sim 2000$) 
with dust corrected $J-K_s$ colors
less than $0.7$ and greater than $1.4$, which were determined to be
non-extragalactic extended sources \citep[see][]{mmkw:03}.
Finally, selecting in the $K_s$-band minimizes 
the inclusion of false sources caused by infrared airglow.

The XSC meets the original 2MASS science requirements: greater than
90\% completeness for extended sources (extragalactic and Galactic) 
with $K_s<13.5$, and free from stellar confusion for $|b|>20\arcdeg$
\citep{jarr:00a}.  In practice, the 2MASS completeness
is a measure of the fraction
of extended sources of a given magnitude that are actually detected.
Huchra \& Mader (2000, private
communication)\footnote{\url{cfa-www.harvard.edu/~huchra/2mass/verify.htm}}
show that the XSC is $99\%$ complete for 
$|b|>30\arcdeg$ sources in the range $12.0 \lesssim K_s \lesssim 13.7$ mag,
using $\log{N}$--$\log{S}$ and $V/V_{\rm max}$ tests.  At magnitudes
brighter than $K_s=11.5$, they determine that greater than $95\%$ of the
known galaxies in \citet{zwicky:68} are found in the XSC.
\citet{jarr:00b} find the XSC completeness remains at $95\%$ for
$K_s\le13.0$ sources well into the Galactic plane ($|b|>5\arcdeg$).
Additionally, \citet{bell:03} cross correlate the XSC with the complete
\citep[$>99\%$][]{stra:02}, spectroscopic galaxy sample drawn from the 
SDSS EDR.  \citet{bell:03} find that in the large (414 square degree), 
high latitude ($|b|>30\arcdeg$) region of the EDR, 
the XSC misses only 2.5\% of the known galaxy population
down to $K_s <13.57$ in dust corrected Kron magnitudes.

\begin{figure} [t] 
\vspace{-20pt}
\centerline{\epsfig{file=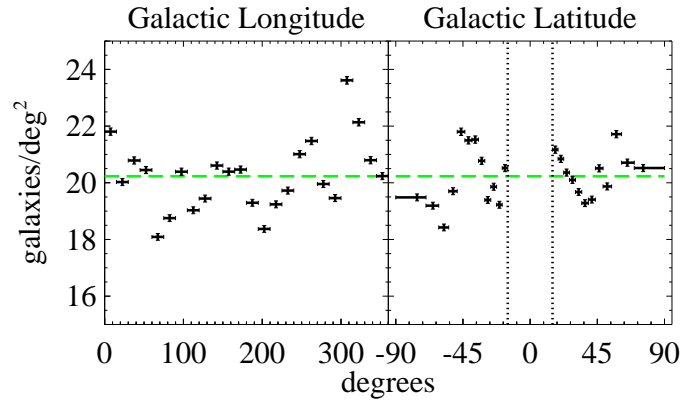, width=\linewidth}}
\caption{ The density of 2MASS galaxies on the sky in bins of Galactic
longitude ($l$), and latitude ($b$).  The horizontal error bars represent the
width of each bin, while the Poisson uncertainties are given by vertical 
error bars.  The dotted lines in the right-hand panel delineate the 
$|b| < 15 \arcdeg$ region corresponding to the Galactic plane.  The 
dashed line shows the average galaxy density of $20.2$ deg$^{-2}$.
There is a significant variation between 
regions but there is no trend with longitude or latitude.
}\label{fig:strip}
\end{figure}

2MASS employs \citet{kron:80} magnitudes to obtain the total flux of
each galaxy.  The 2MASS-defined Kron magnitudes (k\_m\_fe in the XSC)
are elliptical aperture magnitudes with semi-major axes equal to 2.5
times the first moment of the brightness distributions for each
source.  The first moment calculation, the Kron {\it radius}, is
computed to a radius that is five times the 20 mag arcsec$^{-2}$
isophote.  2MASS limits the Kron radius to a $5\arcsec$ minimum (owing
to the PSF; see the 2MASS Explanatory Supplement for details).
Because of the short exposure times \citep[7.8 seconds with a 1.3-m
telescope,][]{skrut:97} of 2MASS, the Kron magnitudes underestimate
the true total flux systematically by approximately $0.1$ magnitudes
\citep{bell:03}.  At a magnitude limit slightly fainter than the 2MASS
science requirement, $K_s\sim13.57$, the XSC galaxy completeness is
quite good, primarily missing\footnote{Some of these missed distant
objects can be found in the 2MASS point source catalog.} only low
surface brightness ($1.5\%$) and distant ($1.0\%$) galaxies both of which
are near the Kron $K_s =13.57$ limit \citep{bell:03}.  We adopt Kron 
magnitudes for this paper and henceforth all cited magnitudes will be Kron
magnitudes.  The small fraction of low surface brightness galaxies
missed in the XSC confirms the expectation of \citet{jarr:00a}.
Moreover, the automated processing of the XSC produces systematically
incomplete photometry for galaxies larger than $50\arcsec$ due to the
typical 2MASS scan width of $8.5\arcmin$ \citep{jarr:00a}.  We include
the 2MASS Large Galaxy Atlas sample (540 galaxies identified in the
XSC database with cc\_flag=`Z') from \citep{jarr:03}, which has been
assembled to account for most of the scan size photometric
incompleteness.

We select initially all $K_s<14$ mag objects from the XSC with $n_{\rm
star}< 10^4 \deg2$.  We apply Galactic foreground dust corrections
from \citet{sfd:98} to each galaxy in the $K_s<14$ catalog and cut at
$K_s=13.57$ mag, resulting in $775,562$ galaxies.  The dust correction
increases the number of galaxies brighter than $K_s=13.57$ by 10\%.
We plot the galaxy density as a function of longitude and latitude in
Figure \ref{fig:strip}, illustrating that the dust-corrected number
densities are fairly uniform for $|b| > 15$.  Ultimately, our final
sample selection includes $501,578$ galaxies with $K_s<13.57$ mag,
$|b|>20\arcdeg$, and Galactic dust corrections less than $0.05$
magnitudes in $K_s$-band (as described in \S 3.2).  We summarize these
sample cuts leading to our final selection in Table \ref{tab:sel}.

\begin{table}[b]
\begin{center}
\begin{tabular}{ccc}
\hline
$N$ & $K_s$ & $|b|(\arcdeg)$\\
\hline
\hline
1,295,895 &   $<14.0$   &  $>0$  \\
  775,562 &   $<13.57$  &  $>0$  \\
  688,464 &   $<13.57$  &  $>10$ \\
  618,333 &   $<13.57$  &  $>15$ \\
  548,353 &   $<13.57$  &  $>20$ \\
  480,794 &   $<13.57$  &  $>25$ \\
  501,578 &   $<13.57$\tablenotemark{a} &  $>20$ \\
\hline
\end{tabular}
\tablenotetext{a}{Only sources with extinction correction $<0.05$ mag.}
\caption{Galaxy Sample Selection.  The number of galaxies in our 2MASS 
catalog is given as a function of $K_s$ magnitude cut and Galactic latitude 
cut.  See \S\ref{sec:cross}.
}\label{tab:sel}
\end{center}
\end{table}

The redshifts of many 2MASS galaxies have been measured as part of other
redshift surveys.  Large redshift samples provide the redshift distribution 
$dN/dz$ of 2MASS galaxies, and
allow calculations of the NIR luminosity function 
\citep{koch:01,cole:01,bell:03}.  For computing
the three-dimensional power spectrum (see \S \ref{sec:3D}), 
we will use redshifts from the SDSS EDR \citep{stou:02} and the
2dF Galaxy Redshift Survey (2dFGRS) 100k Data Release \citep{coll:01}.

\begin{figure} [t] 
\centerline{\epsfig{file=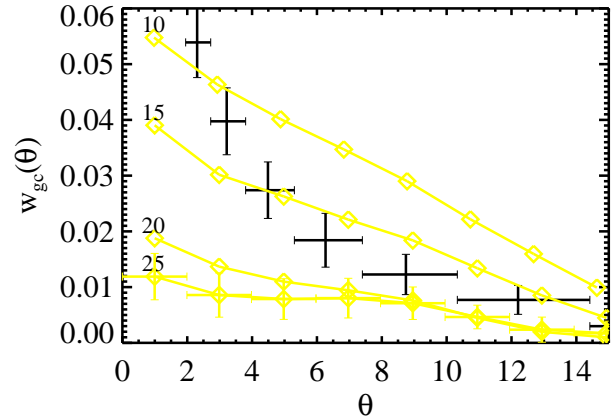, width=\linewidth}}
\caption{Galaxy-star cross-correlations (open diamonds)
for cuts in Galactic latitude of 
$|b| > 10\arcdeg, 15\arcdeg, 20\arcdeg$ and $25\arcdeg$.
Error bars displayed for the $25\arcdeg$ cut are representative 
of the other cuts.  The auto-correlation is shown as naked error bars
for comparison.  We require a latitude cut of 
$|b|>20\arcdeg$ to be free of stellar contaminants out to 
separation angles of $\theta < 15\arcdeg$.
}\label{fig:star}
\end{figure}
  
\section{Cross-Correlation Functions} 
\label{sec:cross}
Following \citet[hereafter Sc02]{scra:02} we analyze the
cross correlation between the galaxy number counts and possible
sources of systematic error.  This is an effective method for
identifying errors, quantifying their magnitude, and determining the
selection limits required to reduce them to manageable levels.  Below
we analyze the cross-correlation signal for four possible
contaminants: stars, dust, seeing, and sky brightness.

To perform the cross correlation, we pixelize the galaxy sample on a
regular grid to determine local galaxy densities.  
We use the HEALPix \footnote{http://www.eso.org/science/healpix/} software
package \citep{ghw:99} to create equal area pixels on the sky.  We use a 
Nside parameter of $64$ generating $N_c = 49,152$ cells each with an area of 
$0.84 {\rm deg}^{2}$.  This yields an average of $15.6$ galaxies per cell so 
very few cells contain no galaxies.
We calculate the average value of each contaminant for a cell using 
the individual measurements for each galaxy within the cell.  
When making a latitude cut, we include all cells
whose centers are above the given latitude including galaxies that
might be below the cut.  To calculate errors, we subsample the data,
dividing the sky into $48$ tiles, with each tile containing $1024$ cells, 
making them roughly $29.3\arcdeg$ on a side. 
Because we will perform a latitude cut and $8$ of the tiles lie along the 
galactic plane we will end up only using $40$ tiles for our analysis.
For each tile we calculate the fractional galaxy and contaminant over 
density in a cell
$i$ by \beq \delta^g_i &=& \frac{n^g_i - \bar{n}^g}{\bar{n}^g} \\
\nonumber \delta^c_i &=& \frac{x^c_i - \bar{x}^c}{\bar{x}^c},
\label{eq:delta}
\eeq 
where the averages are calculated for each tile.
The cross-correlation function, $w_{gc}(\theta)$, is then
\beq
w_{gc}(\theta_\alpha) = 
\frac{\sum_{i,j=1}^{N_c} \delta^g_i \delta^c_j \Theta^\alpha_{ij}}
{\sum_{i^*,j^*=1}^{N_c}\Theta^\alpha_{i^*j^*}},
\eeq
where $\Theta^\alpha_{ij}$ is unity if the separation between cells $i$ and $j$
is within angular bin $\theta_\alpha$ and zero otherwise.  
The error on the mean can be computed by
\begin{equation}
\left[\Delta\bar{w}_{gc}(\theta)\right]^2 = \frac{1}{N_{t}^2}\sum^{N_t}_{i=1}
\left[\bar{w}_{gc}(\theta) - w_{gc,i}(\theta)\right]^2, \label{eq:sub-error}
\end{equation}
where $N_t=40$ is the total number of tiles.  We use this
cross-correlation function to quantify the systematic contributions of
contaminants in the subsections below.

\begin{figure} [t] 
\centerline{\epsfig{file=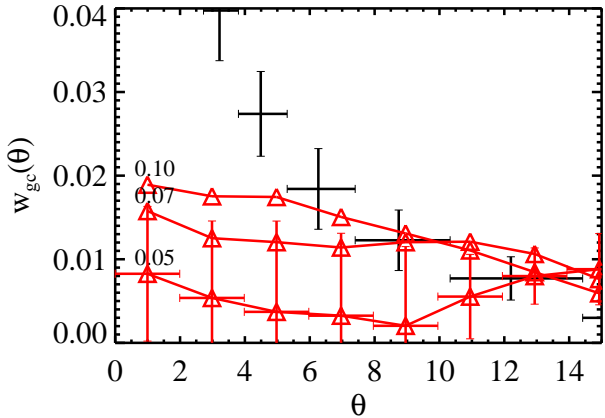, width=\linewidth}}
\caption{Galaxy-extinction cross-correlations (open triangles) 
for three cuts
in the maximum allowed value for the extinction given by
$A(K_s)=0.367\times E(B-V)$, with $E(B-V)$ from \citet{sfd:98}.
Error bars are shown only for the lowest cut and are representative.
The auto-correlation is shown as naked error bars for comparison.
A cut of $A(K_s)\le 0.05$ mag
reduces the cross-correlation signal to less then the 
auto-correlation for $\theta < 10\arcdeg$.
}\label{fig:dust}
\end{figure}

\subsection{Stars}
\label{sec:stars}
During the 2MASS pipeline 
processing, the stellar density $n_{\rm star}$---the number of 
$K_s<14$ mag stars per square degree---is determined for each extended source
on a co-add which is $8.5\arcmin \times 16 \arcmin$ in size.
In regions of high stellar density, stars may be mistakenly
identified as extended sources.  For $95\%$ of the sky, $n_{\rm star}$ 
provides an excellent measure of confusion noise in the XSC.  The stellar 
density saturates in 2MASS at values of 
$n_{\rm star} > 3 \times 10^4 \deg2$ 
\citep{jarr:00b}, occurring near the Galactic center 
($|b|<5\arcdeg$, $350\arcdeg < l < 10\arcdeg$).  Recall that for our galaxy
sample we make a more conservative stellar density cut of 
$n_{\rm star} < 10^4 \deg2$.

To examine the importance of this contaminant we look at the 
cross-correlation of the galaxy density and the stellar density.  
Because $n_{\rm star}$ is a strong function of
Galactic latitude we measure the cross-correlation for different Galactic
latitude cuts and present the results in Figure \ref{fig:star}.  
Including XSC objects with low latitudes ($|b|<15\arcdeg$) results in 
a higher amplitude galaxy-star cross-correlation at all angular scales than
the galaxy auto-correlation. 
That this cross-correlation has a different slope than the galaxy 
auto-correlation suggests that the signal is coming from the 
stellar auto-correlation as multiple star systems are mistakenly identified
as galaxies.
However, if the latitude cut is increased
to $|b|>20\arcdeg$, then the auto-correlation signal is not
contaminated significantly by stars at angular separations of 
$\theta < 15\arcdeg$. Thus, we adopt a $|b|>20\arcdeg$ cut throughout 
this paper.  Making this cut reduces the area of the survey to $67\%$ of
the sky or $28,960$ square degrees. $548,353$ galaxies survive this 
latitude cut.

\begin{figure} [t] 
\centerline{\epsfig{file=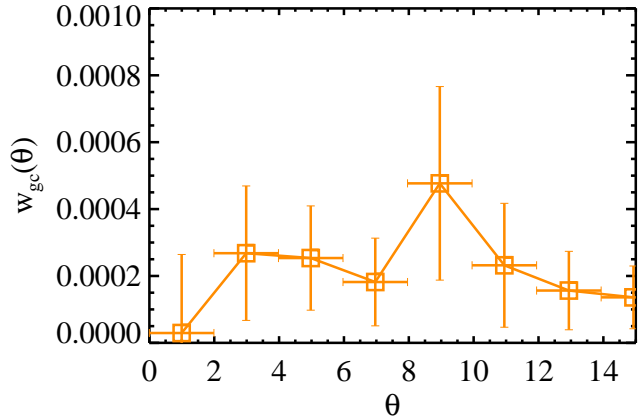, width=\linewidth}}
\caption{The galaxy-seeing cross-correlation (open squares) 
as a function of angle.  The auto-correlation can not be
plotted because it is $>30$ times larger.
Seeing variations are not a source of systematic error.  
}\label{fig:seing}
\end{figure}

\subsection{Dust}
Though we have attempted to correct our galaxy sample for the effects of
foreground extinction using \citet{sfd:98}, dust may remain
an important contaminant.  There is a strong correlation 
between stellar density and dust extinction, both of which increase 
substantially towards the Galactic plane.  Here we look at the 
cross-correlation between dust extinction and galaxy density only at 
$|b| > 20\arcdeg$, owing to the stellar contamination at lower latitudes.
In Figure \ref{fig:dust}, we show the 
galaxy-extinction cross-correlation for three cuts in magnitudes of 
$K_s$-band dust extinction.  
To bring the galaxy-extinction cross-correlation below the auto-correlation
we mask out those cells with $K_s$-band extinction of
$A(K_s)=0.367\times E(B-V)>0.05$, which amounts to $2,602$ cells.
This leaves and area of $23,345$ sq. degrees and $501,578$ galaxies. 

Our use of a dust correction to galaxy magnitudes does increase the
dust cross-correlation.  For example, the amplitude of the
galaxy-extinction cross-correlation is reduced by about a third for a
catalog {\em without} a dust correction.  However, the dust correction
increases the number of galaxies by 10\%, and therefore the error
introduced by the correction is a minor contribution to the systematic
error budget.

\begin{figure} [t] 
\centerline{\epsfig{file=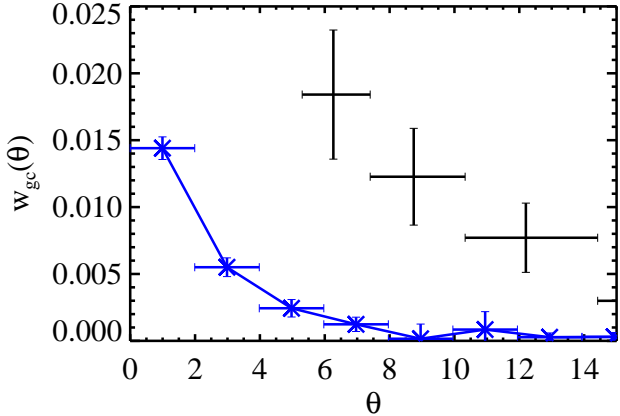, width=\linewidth}}
\caption{The galaxy-background (Xs) cross-correlation function
as a function of angle.  The background error correlates with galaxy
density as expected, yet 
it is always significantly less than the auto-correlation (naked error bars).
Sky background fluctuations are a negligible source of systematic error.
}\label{fig:back}
\end{figure}

\subsection{Seeing}
Great care has gone into ensuring
that the final 2MASS release has a highly uniform seeing over the entire
survey area.  The pipeline processing allowed tracking of
the point-spread function (PSF) as a function of time.\footnote{2MASS mapped
the sky using overlapping scans roughly $8\farcm5$ wide by $6\arcdeg$
long.  The scan direction followed the declination axis, with each scan
covering $6\arcdeg$ in approximately 6 minutes.}
The mean radial size of the 2MASS PSF over a scan could vary over periods 
as short as a few minutes. 
Depending on the local density of stars, the seeing was tracked on
time-scales ranging from 2---30 seconds.  Scans with
poor seeing were reobserved under better conditions to maintain the seeing
uniformity.  For each source in the XSC,
the pipeline measured the average radial extent of the PSF in each passband
(called ``ridge shape'' and denoted k\_sh0 in the XSC for the $K_s$-band),
analogous to a full width at half maximum (FWHM) measurement of the varying 
PSF.  The full details of the PSF characterization and seeing
tracking are given in \citep{jarr:00a}.

We use k\_sh0 to test for systematic errors in the angular correlation of
galaxies caused by spatial variations in seeing.
One might expect that observations made under poor seeing conditions
would lead to more spurious galaxy detections, causing
a correlation between seeing and galaxy density. For our sample the
mean seeing is k\_sh0 = 0.992 with 0.049 rms scatter, corresponding to a
typical survey seeing of $2.1\arcsec$ FWHM with 5\% uncertainty.  Given
this uniformity, we do not expect the seeing to be a major source
of systematic errors. 
One might want to check for seeing variations over the scan width 
of $8.5\arcmin$. However, even in the densest regions there are only
a few 2MASS galaxies in such a small area so a cross-correlation 
cannot be performed.  
In Figure \ref{fig:seing} we plot the
galaxy-seeing cross-correlation, whose amplitude
is more than $30$ times less than the auto-correlation amplitude.  
Hence, seeing variations are a negligible source of systematic error.

\subsection{Sky Background} 
In $K_s$-band, the dominant source of background flux
is thermal continuum emission from the atmosphere.
While the $K_s$ background is less
severe and variable than the airglow induced $J$ and $H$-band 
backgrounds, the sky brightness at $K_s$ is still quite high
at $13.3\pm0.3$~mag arcsec$^{-2}$.  This is orders of magnitude brighter than
the typical outer isophotes of 2MASS galaxies, which have
$\mu_{K}>19$~mag arcsec$^{-2}$.  Furthermore, the $K_s$
background can be variable and may still produce
high-frequency features extending to tens of arcseconds.  To mitigate
these features, the 2MASS 
processing includes a sophisticated background fitting and removal scheme
that is described fully in \citet{jarr:00a}.  

The rms background error caused by Poisson noise, which
depends on the variable sky brightness, might also produce sources of
systematic error in calculating the angular correlation of 2MASS galaxies.
The median background local to each source is determined in each passband 
and given in the XSC database (k\_back for $K_s$).  For each source in our 
sample, we estimate the background error from co-added and resampled
images
\begin{equation}
\sigma_{\rm bkg}^2 = \frac{k_{\rm f}^2}{6n_{\rm r}^2} \left[ \frac{n_{\rm r}({\rm k\_back})}{G} + \frac{\sigma_{\rm RN}^2}{G^2} \right] .
\end{equation}
The factor of 6 represents the number of co-added frames comprising
each source image.  Each $2\arcsec$ pixel has been resampled ($n_{\rm r}=4$)
into
$1\arcsec$ pixels and smoothed with a kernel giving rise to the factor
of $k_{\rm f}=0.5853$.  In $K_s$-band 2MASS has an average gain of
$G=8.2$ electrons/DN and an average read noise of
$\sigma_{\rm RN}=52$ electrons (T. Jarrett, 2001, private communication)
\footnote{spider.ipac.caltech.edu/staff/jarrett/2mass/3chan/noise/index2}.

\begin{figure} [b] 
\centerline{\epsfig{file=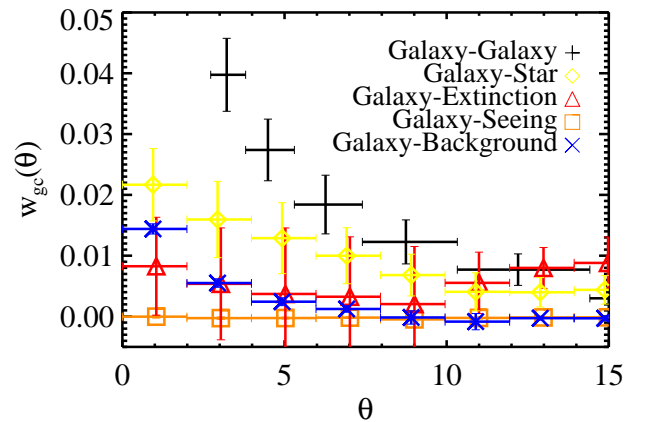, width=\linewidth}}
\caption{ The cross-correlations with all four investigated
contaminants using our cuts in Galactic latitude of 
$b > 20\arcdeg$ and in extinction of $K_s < 0.05$ compared to
the galaxy auto-correlation signal. On 
angular scales
$<10\arcdeg$ the auto-correlation amplitude is significantly higher then
the cross-correlation with any of the contaminants. 
}\label{fig:cross}
\end{figure}

A larger sky uncertainty increases the difficulty of galaxy detection
and leads to a correlation between background error and 
galaxy density.  We find an average k\_back of 475 DN with scatter 125 DN, 
corresponding to a background uncertainty of $20.02\pm0.13$~mag arcsec$^{-2}$.
This agrees with the typical $1\sigma$ error of 20.0~mag arcsec$^{-2}$ in $K_s$
\citep{jarr:00a,jarr:03}.  The estimated background error
is quite uniform and the galaxy density and background error correlation is
weak, as we show in Figure \ref{fig:back}.
The cross-correlation amplitude is always 
smaller than that of the galaxy auto-correlation and hence background
fluctuations are a negligible source of error in determining $\acf$.

\subsection{Summary of Systematic Effects} 
Figure \ref{fig:cross} shows the cross-correlations of the four possible 
contaminants in comparison with $\acf$.  With the Galactic latitude cut 
of $|b| > 20\arcdeg$ and the dust extinction cut of $K_s < 0.05$,
we see that the
contaminants are significantly below the auto-correlation signal for 
angular scales $\theta < 10\arcdeg$.
We measure $\acf$ on larger scales, but caution that for $\theta > 10\arcdeg$
systematic errors might be as large as the signal.

\begin{figure} [t] 
\centerline{\epsfig{file=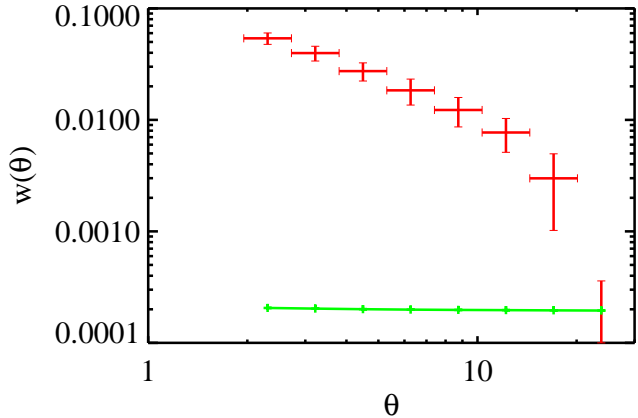, width=\linewidth}}
\caption{The integral constraint correction, $\Delta \hat{w}(\theta)$, compared
to the angular correlation function $\acf$ for the pixelized estimator.  
The integral constraint is small because it is suppressed by a factor 
$N_c^2$ where in this case $N_c=32,512$.  Thus the integral constraint is not 
relevant for large surveys like 2MASS and SDSS except at the largest angular
scales.
}\label{fig:ic}
\end{figure}

\section{The Angular Correlation Function}
\label{sec:acf}
\subsection{Estimators, Biases, and Errors}
Following Sc02 we use two estimators for the angular correlation function. 
The first, which we will refer to as the pair-estimator, compares positions 
of the observed galaxies (data) to the positions of random points by 
counting the number of pairs in an angular bin $\theta_\alpha$ normalized 
by the total number of pairs.  We use one million randomly placed points 
for the comparison, avoiding the regions that were masked as described in \S3.
The angular correlation can then be computed using the 
the now standard estimator introduced by \citet{ls:93},
\beq
\acf= {{\langle DD \rangle -2\langle DR \rangle + \langle RR\rangle} 
\over{\langle RR\rangle}} \,,
\eeq
where DD, DR, and RR are the normalized number of data-data, data-random and 
random-random pairs in an angular bin $\theta$.
At large angular scales counting pairs becomes computationally expensive,
so instead we calculate the galaxy density in a cell as described 
in \S\ref{sec:cross} (eq. \ref{eq:delta}), which we refer to as the 
cell-estimator.  Then the value of the angular correlation function is 
given by  
\beq
\acf = {{\sum \delta^g_i \delta^g_j \Theta^{ij}_{\alpha}} 
\over{\sum_{ik} \Theta^{ij}_{\alpha}}},
\eeq
where again $\Theta^{ij}_{\alpha}$ is unity if the angular separation 
of cell $i$ and cell $j$ is within $\theta_\alpha$ and zero otherwise.
We use the pair-estimator for $\theta < 5\arcdeg$ and the cell-estimator
for $\theta > 2\arcdeg$ which gives three angular bins where we use both
methods to check that they generate the same results.
In the overlapping regions we will use the pair-estimator which 
should be more accurate because the data has not been smoothed.

All estimates of $\acf$ are subject to a statistical bias 
referred to as the ``integral constraint'' 
\citep{peeb:80,bern:94,hg:99}.  This bias arises because the
estimate of the mean number density in a given cell enters into the 
estimator of the angular correlation function nonlinearly. 
The integral constraint correction, $\Delta \hat{w}(\theta)$, for
the cell-estimator can be calculated to be (see Sc02), 
\beq
\Delta \hat{w}(\theta_\beta) = 
\left[1 + (2c_{12} - 3)w(\theta_\beta) 
\right] \frac{1}{N_c^2} \sum_{i,j} w(\theta_{ij}),
\label{eq:bias}
\eeq
where $c_{12}\approx 2$ and the sum is over all cells. Since this value is 
suppressed by a factor
of $N_c^2$ this bias is extremely small for large surveys like 2MASS and 
SDSS. Figure \ref{fig:ic} shows $\Delta \hat{w}(\theta)$ in comparison 
with $\acf$.  One clearly sees that the integral constraint can safely be 
ignored at all angular scales where we can measure $\acf$.

\begin{figure} [t] 
\centerline{\epsfig{file=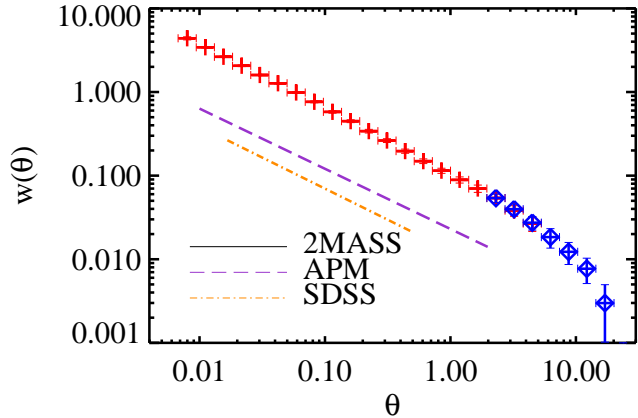, width=\linewidth}}
\caption{ The angular correlation function of 2MASS galaxies.  Bins
evaluated using the pixelized estimator are shown with an X, otherwise
the pairs estimator is used (crosses).  The errors are based on Jackknife
resampling.  The 
best-fit power laws of the APM \citep[long dash;][]{madd:90} and 
SDSS \citep[dot-dash;][]{conn:02} surveys 
are shown for comparison.
The solid line is a power law with a slope of 
$\gam$ and an amplitude at $1\arcdeg$ of $\A$.
}\label{fig:acf}
\end{figure}

To calculate errors we use the Jackknife resampling method, the method
that Sc02 found to be the best data-only method for estimating errors.  
We remove one tile of $1024$ cells and then use the two estimators 
described above to determine $\acf$ on the remaining data set.  
We therefore make $N_t=40$ measurements of the angular correlation function 
in $25$ angular bins. To calculate errors we determine a covariance
matrix out of the $40$ measurements.   
The $\alpha\beta$ element of the covariance matrix is computed by
\beq
C_{\alpha\beta}= 
{{N_t}\over{N_t-1}}\sum_{k=0}^{N_t} 
\left[w_k(\theta_\alpha)-\bar{w}(\theta_\alpha)\right]
\left[w_k(\theta_\beta)-\bar{w}(\theta_\beta)\right]
\eeq
where $w_k(\theta_\alpha)$ refers to the $k$th measurement of the angular 
correlation function on the angular scale $\theta_\alpha$.

Because of the strong covariance between different angular bins, it is 
necessary to include it when fitting a model to the data.
Thus we minimize the complete form of the statistic
\beq
\chi^2 = \sum_{\alpha,\beta} \left[w(\theta_\alpha)-w_m(\theta_\alpha)\right]
C^{-1}_{\alpha\beta} \left[w(\theta_\beta)-w_m(\theta_\beta)\right]
\label{eq:chi}
\eeq
where $w_m(\theta)$ is a model of the angular correlation function
\citep[see][for a discussion]{bern:94}.

\begin{figure} [t] 
\centerline{\epsfig{file=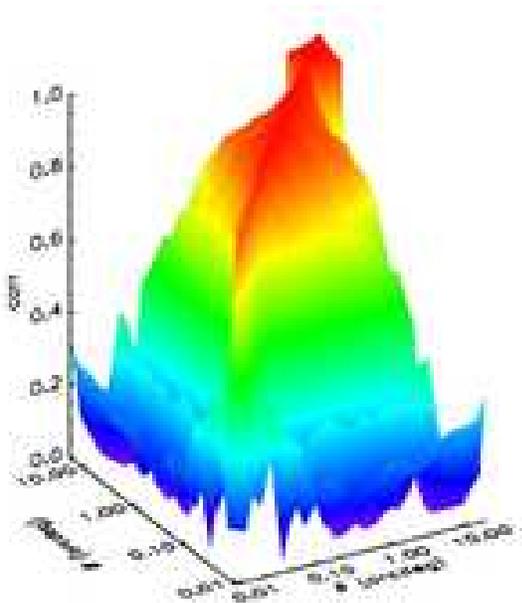, width=\linewidth}}
\caption{The correlation matrix of the angular correlation function of 
2MASS galaxies.  The correlation matrix is the covariance matrix normalized
by its diagonal elements.  One sees that for bins less than $0.06 \arcdeg$
the data is fairly independent.  But at scales larger than this there is 
a strong correlation between different data bins.
}\label{fig:corr}
\end{figure}

\subsection{Results}
The angular correlation function of 2MASS galaxies is shown in Figure 
\ref{fig:acf}.  We fit a power law of form $\acf=A\theta^{1-\gamma}$ out to 
angular scales of $2.5\arcdeg$ to the estimated $\acf$.  Note that the 
uncertainties of the power-law fit are substantially underestimated if one 
ignores the covariance between angular bins.  Hence
to fit a power-law form to our estimated angular correlation function
we use Singular Value Decomposition (see \S\ref{sec:inv}) to remove
the oscillatory modes from the covariance matrix.  We then find a best fit
power law with an amplitude at $1\arcdeg$ of $A=\A$ and $1-\gamma=\gam$.
However, we find that a power law is 
{\it not} a good fit to the angular correlation function 
($\chi^2/{\rm d.o.f.} = 2.4$ using eq. \ref{eq:chi}) 
using the full covariance matrix (see discussion below).
As shown in Table \ref{tab:acf}, the slope
of $\acf$ from 2MASS is 
slightly higher but in overall 
agreement with the slopes found in
APM \citep{madd:90} and SDSS EDR \citep{conn:02}. The power-law fit to
the 2MASS data extends past $\theta=1-2\arcdeg$, where the APM and EDR
angular correlation functions start to deviate from a power law.  This 
is expected since the same physical scale will correspond to a larger
angular scale in the shallower 2MASS catalog. 

\begin{figure} [t] 
\centerline{\epsfig{file=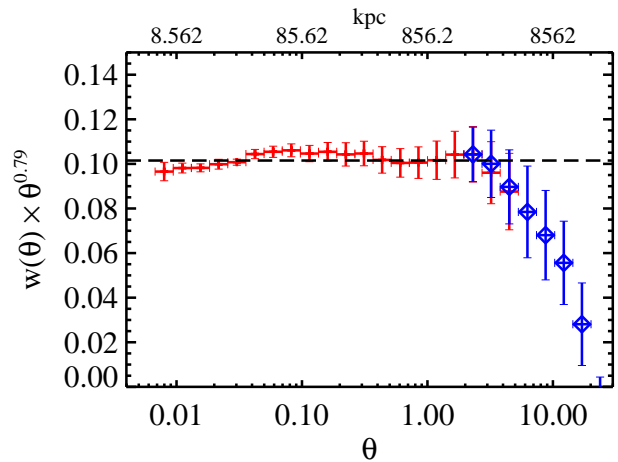, width=\linewidth}}
\caption{ The angular correlation function of 2MASS galaxies multiplied
by $\theta^{0.79}$. Small oscillations around the power-law form are apparent
and are statistically significant.  The horizontal dashed line with amplitude 
$A=0.10$ is the best fit power law. 
Bins evaluated using the pixelized estimator are shown with an X, otherwise
the pairs estimator is used (crosses).
}\label{fig:diff}
\end{figure}

\begin{table*}
\begin{center}
\begin{tabular}{ccrcc}
\hline
Survey & A & 1-$\gamma$ & $z_m$ & sample selection\\
\hline
\hline
APM    & $0.028\pm0.003$  & $-0.70\pm0.03$ & $0.11$  & $b_J < 20  $\\
SDSS   & $0.013\pm0.060$  & $-0.74\pm0.04$ & $0.17$  & $18<r_*<19 $\\
2MASS  & $0.10\pm0.01$   & $-0.79\pm0.02$ & $0.074$ & $K_s < 13.5$\\
2MASS  & $0.08\pm0.01$   & $-0.79\pm0.02$ & $0.080$ & $13.5< K_s < 12.5$\\
2MASS  & $0.22\pm0.02$   & $-0.79\pm0.02$ & $0.047$ & $12.5< K_s < 11.5$\\
2MASS  & $0.48\pm0.03$   & $-0.48\pm0.02$ & $0.029$ & $11.5< K_s < 10.5$\\
2MASS  & $0.15\pm0.01$   & $-0.79\pm0.02$ & $0.071$ & $\mu_K < 17.75$\\
2MASS  & $0.08\pm0.006$  & $-0.76\pm0.02$ & $0.079$ & $\mu_K > 17.75$\\

\hline
\end{tabular}
\caption{The results of fitting the the angular correlation 
function with a functional form $A\theta^{1-\gamma}$ for APM \citep{mes:96}, 
SDSS \citep{conn:02} and this work.  We also show the fitted parameters 
when the 2MASS sample is divided by effective surface brightness
and into magnitude bins.  
}\label{tab:acf}
\end{center}
\end{table*}

The amplitude
of the 2MASS angular correlation function is 3.6 (7.7) times larger than the
amplitude found in the APM (SDSS EDR) surveys.
We expected there would be an amplitude difference both
because 2MASS galaxies are  brighter than either the APM or SDSS EDR survey
and because 2MASS galaxies have a shallower redshift distribution.   Assuming
that the galaxies in all three surveys have the same 3-dimensional power 
spectrum, we can use Limber's equation (eq. \ref{eq:limber})
to calculate that the change in amplitude in $\acf$ when going from the median 
redshift of SDSS ($0.17$) to the median redshift of 2MASS ($0.074$) is
approximately a factor of 5.  From Table \ref{tab:acf} we see that the ratio
of amplitudes is a factor of $7.7$, hence even after including the median
redshift difference the 2MASS $\acf$ amplitude is still $50\%$ higher than 
SDSS.  We attribute the remaining amplitude offset to the difference in 
luminosity of the galaxies being sampled.  The average ($r-K_s)$ color of a 
galaxy is approximately $3.0$ so our magnitude cut of $K_s < 13.57$ 
corresponds to $r < 16.5$.  Therefore, the faintest 2MASS galaxies are two 
magnitudes brighter than the mean of the brightest magnitude bin used in the 
SDSS study.  More luminous galaxies are more clustered than less luminous ones
\citep{norb:01,zehavi:02}.

\begin{figure} [b] 
\centerline{\epsfig{file=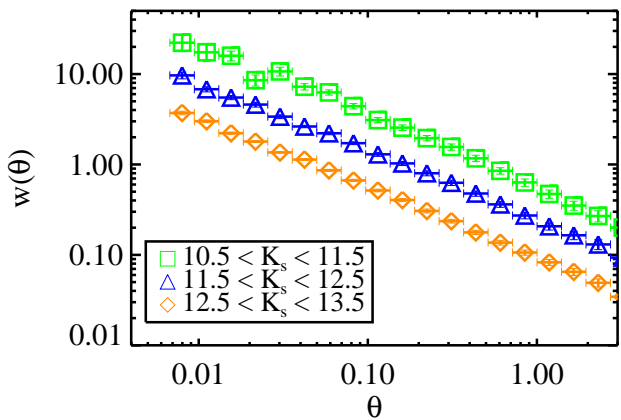, width=\linewidth}}
\caption{The angular correlation functions of 2MASS galaxies divided 
into bins of $\Delta K_s=1$ centered on $K_s$=11,12,13.
The error bars, which are smaller then the symbols, decrease with $K_s$
magnitude since there are fewer bright galaxies.
The slopes remain the same while the
amplitudes increase for brighter galaxies as expected.
}\label{fig:acfmag}
\end{figure}

In Figure \ref{fig:diff}, we plot the angular correlation function of
2MASS galaxies multiplied by $\theta^{0.79}$ to examine the deviations
from a pure power law.  The poor value of $\chi^2/{\rm d.o.f.}$ for the
power-law fit, the straight-horizontal line, may seem surprising since the
magnitude of the deviations are similar to the error bars.  However,
the depicted error bars follow from the diagonal elements of the
correlation matrix and do not describe independent variance.  Owing to
significant off-diagonal terms in the covariance matrix, the
displacement of adjacent $w(\theta)$ bins towards the power law fit
gives a very large contribution to $\chi^2$.  
The correlation matrix of $w(\theta)$ is shown in Figure \ref{fig:corr}.
The importance of off diagonal elements is clearly seen in the figure.
Thus, the oscillatory
behavior scene in Figure \ref{fig:diff} is significant.  These may be
the result of wiggles in the power spectrum caused by baryon
oscillations as recently detected in the 2dFGRS power spectrum
\citep{perc:01}.  The oscillations seen here, however, are at much
smaller scales then those seen in the 2dFGRS and are in the nonlinear
regime.  Such oscillations are expected in
halo occupation distribution descriptions of galaxy clustering
\citep{selj:00,bw:02,zehavi:04} and the dip centered around 0.8 degrees,
corresponding to $700h^{-1}$ kpc at the median redshift of our survey, occurs
at about the expected physical scale.

\begin{figure} [b] 
\centerline{\epsfig{file=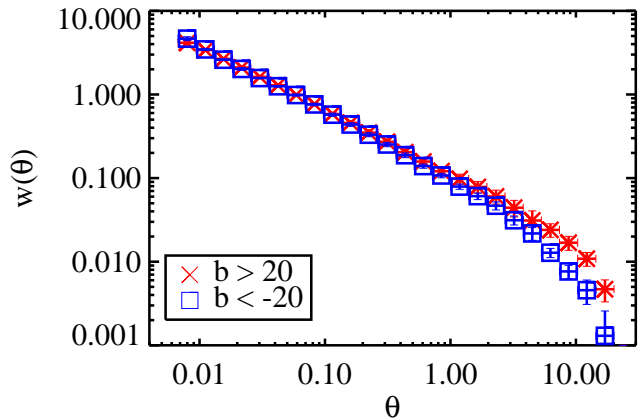, width=\linewidth}}
\caption{The angular correlation function of 2MASS galaxies for
northern Galactic latitudes ($b > 20\arcdeg$ diamonds) and southern 
Galactic latitudes ($b < -20\arcdeg$ squares). For angular scales less 
then $0.5\arcdeg$ the two hemisphere yield consistent results. However,
at larger angular scales there is stronger clustering in the northern
Galactic hemisphere.  
}\label{fig:acfns}
\end{figure}

We also calculate the angular correlation of 2MASS galaxies split 
into bins of $\Delta K_s=1$ centered on $K_s$=11, 12, 13 and
plot each $\acf$ in Figure \ref{fig:acfmag}.
The error bars at a given angular separation increase
for brighter magnitudes owing to their lower numbers.
Limber's equation \citep{limber:53} implies that apparently brighter galaxies
should have parallel $\acf$ with higher amplitudes owing to their closer
redshift distribution.  Using the median
redshift of the galaxies in each magnitude bin (see Table \ref{tab:acf})
we estimate that the amplitude ratio should be $6:2.5:1$, going
from the brightest
to the faintest bin.  In Table \ref{tab:acf} we present the parameters of the
power-law fits to $\acf$ for each magnitude bin and
confirm that the differences in amplitude can be fully explained by differences
in median redshift, consistent with true clustering.

Because 2MASS is an all sky survey with 1\% photometric uniformity on large
angular scales
\citep{niko:00}, we have two independent volumes, one for northern 
$b > 20\arcdeg$ and one for southern $b < -20\arcdeg$ Galactic latitudes.  
We can compute $\acf$ independently
for the two hemispheres as a check on our procedure and to examine the 
effects of cosmic variance, plotted in Figure \ref{fig:acfns}.
At angles less than $0.5\arcdeg$ there is good agreement.  However, at larger
angles the northern Galactic hemisphere shows more clustering, caused by real
differences in the observed large scale structure between the northern and
southern Galactic hemispheres \citep[][Fig. 2]{mmkw:03}.
For small angular scales, however, this cosmic variance
is unimportant.

\begin{figure} [t] 
\centerline{\epsfig{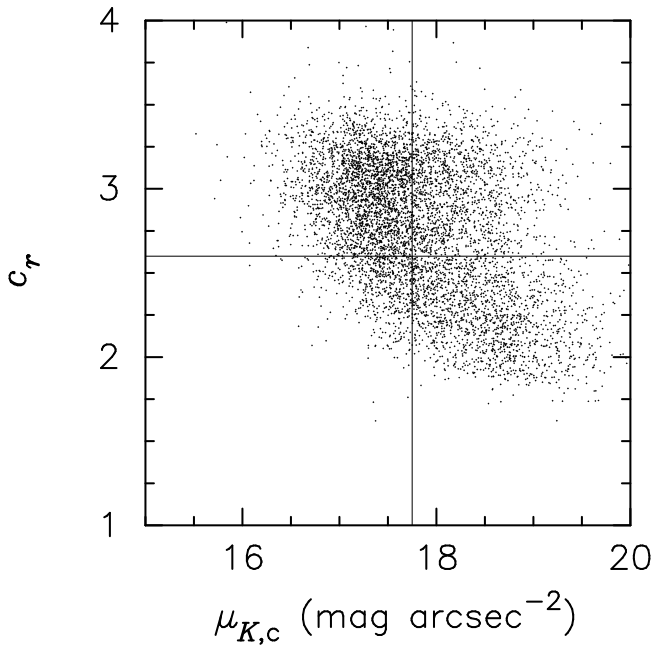}}
\caption{SDSS $r$-band concentration $c_r = r_{90}/r_{50}$ 
versus effective surface brightness 
$\mu_K = K_s + 2.5\log_{10}{(2\pi r^2_e)}$.
The horizontal dotted line indicates $c_r=2.6$, chosen by SDSS to divide
early from late-type galaxies.  The vertical dashed line at 
$\mu_K<17.75$ mag arcsec$^{-2}$ shows our adopted morphology divider.
}\label{fig:csb}
\end{figure}

One would like to investigate the dependence of $\acf$ on 
galaxy morphology. In the optical, galaxy morphology can be estimated 
using either galaxy color or concentration. In the infrared, however,
the ($J-K_s$) colors contain little information about the galaxy type, 
since all galaxies have ($J-K_s$) colors that are tightly peaked around
1.0.  Furthermore, \citet{bell:03} demonstrate that
the 2MASS concentration measurement is not very accurate for faint ($K_s>12$)
galaxies.  Instead, as a measure of galaxy morphology we use effective surface
brightness 
\beq
\mu_K = K_s + 2.5\log_{10}{(2\pi r^2_e)},
\eeq
where $r_e$ is the half-light radius of the galaxy, and $K_s$ is the 
extinction-corrected Kron magnitude.
For $6,238$ 2MASS galaxies also in the SDSS EDR we show in Figure \ref{fig:csb}
that $\mu_K$ is 
correlated with the SDSS optical concentration $c_r = r_{90}/r_{50}$,
where $r_{90}$ and $r_{50}$ are the radii within which $90\%$ and
$50\%$ of the galaxy flux are contained, respectively.
The SDSS collaboration has adopted $c_r=2.6$ to separate between early and 
late types in a rudimentary fashion
\citep[e.g.][]{stra:01,kauf:03}. 
Using $c_r\ge2.6$, we find that 2630/3203 (82\%) of the high surface brightness
($\mu_K<17.75$ mag arcsec$^{-2}$) galaxies are early-type.

\begin{figure} [t] 
\centerline{\epsfig{file=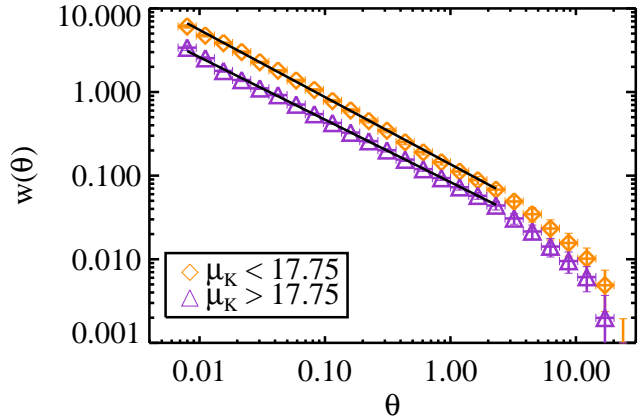, width=\linewidth}}
\caption{The angular correlation function of high and low surface
brightness 2MASS galaxies.
Galaxies with $\mu_K < 17.75$ (diamonds) are
more clustered and have a slightly steeper slope than galaxies with
$\mu_K > 17.75$ (triangles).  
}\label{fig:sb}
\end{figure}

Dividing the sample at the median $\mu_K = 17.75$, we plot
the angular correlation functions of the two populations in Figure 
\ref{fig:sb}.  The higher surface brightness galaxies are more 
clustered with an amplitude at $1\arcdeg$ almost twice that of the 
lower surface brightness galaxies (see Table \ref{tab:acf}).  Unlike the
difference in amplitude for the different magnitude bins, which are caused
by differences in the median redshifts, our high and low surface brightness
galaxies have nearly identical median redshifts of 0.079 and 0.071, 
respectively.  Hence, the higher $\mu_K$ (early-type) galaxies truly are 
more clustered in three dimensions than the low $\mu_K$ (late-type) galaxies 
in our sample. This increased clustering amplitude for early type galaxies 
has also been seen in the EDR three-dimensional correlation function 
\citep{zehavi:02} and is related to the morphology-density relation 
\citep{dres:80}.

The slopes of the two angular correlation functions are marginally
different; they disagree at the $90\%$ confidence level. Such a disagreement
could result from nonlocal biasing or nonlocal causes for the 
morphology-density relation \citep{nbw:00,sw:98}.  It would be interesting 
to see if any galaxy formation scenario matches this slight disagreement.

\section{The 3D Power Spectrum}
\label{sec:3D}
In this section we invert the estimated angular correlation function to 
measure the three-dimensional power spectrum.  The relationship between 
$\acf$ and $P(k)$ can be expressed as 
\beq
\label{eq:limber}
\acf=\int^{\infty}_0 dk k P(k)g(k\theta)
\eeq
where the kernel $g(k\theta)$ is given by 
\beq
\label{eq:kernel}
g(k\theta)= {{1}\over{2\pi}}\int^{\infty}_0 
dz J_0(k\theta X(z)) (\frac{dN}{dz})^2{{dz}\over{dX}} F(z)
\eeq
\citep{limber:53,be:93}. In the above equation $X(z)$ is the comoving 
distance to a redshift $z$, $J_0$ is the zero order Bessel function and 
$F(z)$ is a function that describes the redshift evolution of density 
fluctuations and any evolution of the galaxy population.  Following 
\citet{dode:02} we set $F(z)=1$, which is a much better approximation for 
the very local 2MASS data than for other surveys. $\frac{dN}{dz}$ is the
probability distribution of galaxy redshifts in the survey, i.e. 
the number of galaxies per redshift bin normalized to unity.  
The comoving distance is
\beq
X(z)={{c}\over{H_0}}\int^z_0 {{dz'}\over{E(z')}}
\eeq
where   
\beq
E(z)&=&{{dz}\over{dX}}\\
    &=&\sqrt{\Omega_m(1+z)^3+\Omega_K(1+z)^2+\Omega_{\Lambda}} \nonumber
\eeq
\citep{peeb:80}. Thus, with the measured angular correlation function and 
knowledge of the redshift distribution of the galaxies in the survey, it
is possible to estimate $P(k)$.  We adopt the currently fashionable
$\Lambda$CDM model \citep{sper:03}, $\Omega_{m}=0.3$ and 
$\Omega_{\Lambda}=0.7$, and note that the depth of 2MASS makes cosmological 
dependence small. 

There are a number of methods for performing the inversion.
We will use the method of Singular Value Decomposition advocated by 
\citet{ez:01}, and used for the SDSS analysis of the EDR \citep{dode:02}

\subsection{Redshift Distribution}
\begin{figure} [t] 
\centerline{\epsfig{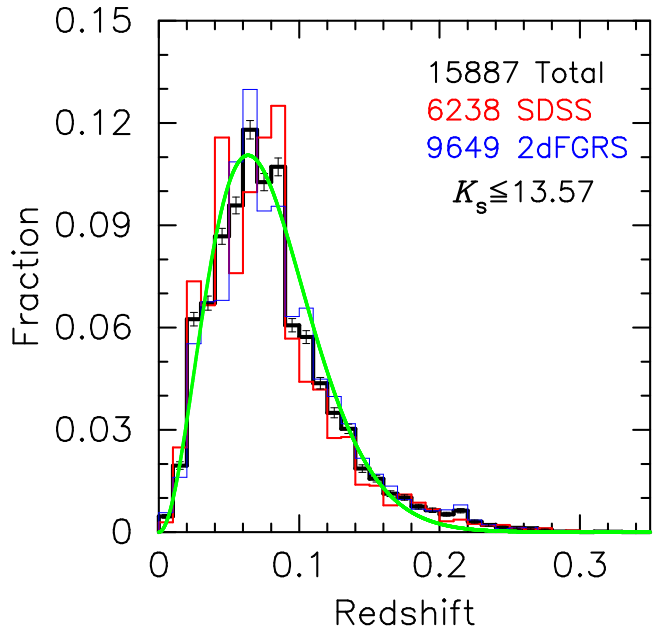}}
\caption{Relative redshift distributions of 2MASS galaxies identified in
the SDSS EDR (red, $N=6238$) and the 2dFGRS 100k release (blue, $N=9649$).
The total distribution for the $15,887$ sources is given in bold with 
Poisson errors shown for each bin.  The smooth curve is our best fit to 
equation (15), which has median redshift $z_m=0.0741$.
}\label{fig:dndz}
\end{figure}

To determine the redshift distribution of 2MASS galaxies we use 
redshifts measured by the SDSS EDR \citep{stou:02} and by the 
2dFGRS 100k release \citep{coll:01}.  We identify $6,238$ 2MASS galaxies with 
redshifts in the SDSS EDR and $9,649$ 2MASS galaxies with redshifts in the
2dFGRS 100k release.  The distribution of redshifts from the two samples
(shown in Figure \ref{fig:dndz}) are similar, although statistically
significant differences, which are caused by the local
large-scale structure, are apparent at most redshifts.

\begin{figure} [t] 
\centerline{\epsfig{file=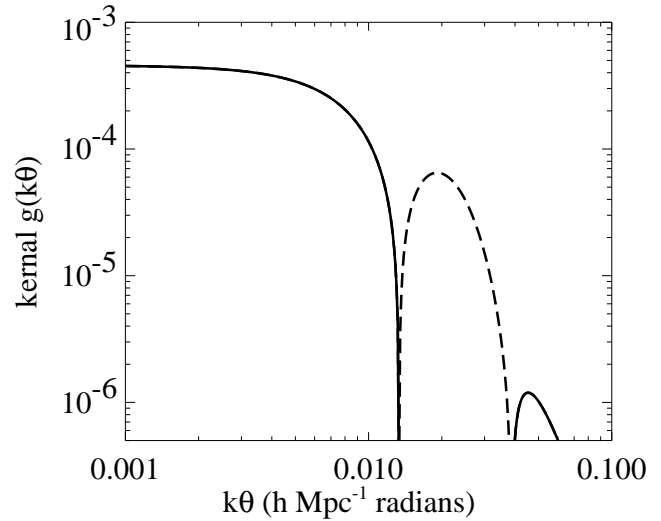, width=\linewidth}}
\caption{ The absolute value of the kernel as a function of $k\theta$.  
The dashed line shows where the kernel is negative.  The inversion 
will be most sensitive to the first zero of the kernel,
which occurs at $k\theta =0.013$.
}\label{fig:kernel}
\end{figure}

For the combined sample of $15,887$ redshifts we fit the functional form
\beq
\label{eq:pz}
\frac{dN}{dz}={{3z^2}\over{2(z_m/1.412)^3}}\exp{[-(1.412z/z_m)^{3/2}]}
\eeq
\citep{be:93}, where $z_m$ is the median redshift. We use 
bootstrap resampling to estimate the uncertainty in our determination of
$z_m$ and find that $z_m=0.0741\pm0.004$.  This statistical uncertainty 
is probably a slight underestimate of the cosmic variance uncertainty 
as evidenced from the median redshifts of the two subsamples, which is
$0.0746$ and $0.0735$ for 2dFGRS and SDSS, respectively. These are both more 
than one sigma from the median of the combined sample. 
We compare the functional form $dN/dz$
to the data in Figure \ref{fig:dndz}. The kernel resulting from using this 
$dN/dz$ in equation \ref{eq:kernel} is shown in 
Figure \ref{fig:kernel}.

\subsection{Inverting the Angular Correlation Function}
\label{sec:inv}
In practice we measure the angular correlation function in $N_{\theta}$ bins
and would like to estimate the power spectrum in $N_k$ bins.  We can 
transform the integral in equation (\ref{eq:limber}) into
a discrete sum and write the equation in matrix form as
\beq
w=GP
\eeq
where $w$ is an $N_{\theta}$ element vector containing the value of $\acf$ in
each bin, $P$ is an $N_k$ element vector containing the values of $P(k)$ in 
each bin, and the $N_{\theta} \times N_k$ matrix $G$ is a discretization 
of equation (\ref{eq:limber}).  

The inversion is thus reduced to a standard problem in linear algebra; the
best fit power spectrum being given by $P=G^{-1}w$ and the power spectrum
covariance matrix by $C_P=G^tC_wG$. Singular Value (SV)
Decomposition \citep[for a review see][ \S2.6 and \S15.6]{press:92} provides
an effective method of solving and analyzing this problem. The matrix $G$ 
can be decomposed into 
$G=UWV^t$ where $W$ is a square diagonal matrix of the singular values and 
$U$ and $V$ are unitary. 
Then $G^{-1}$ is simply given by $G^{-1}=VW^{-1}U^t$. However, those 
elements of $W$ that are singular or very small will make a large 
contribution to $G^{-1}$. To remove this contribution, these eigenvalues 
are set to zero in the (pseudo-) inverse matrix $W^{-1}$.  In practice, 
we remove eigenvalues that cause large oscillations in the resulting power 
spectrum.

\begin{figure} [t] 
\centerline{\epsfig{file=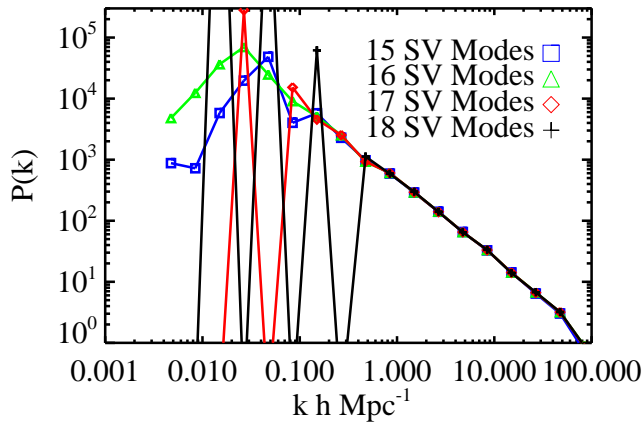, width=\linewidth}}
\caption{The recovered power spectrum for different choices in the 
number of SV modes retained for the inversion.  Retaining too many modes
leads to wild oscillations in $P(k)$.  With $16$ SV modes we are sensitive
to small $k$ but do not exhibit oscillations.  
}\label{fig:modes}
\end{figure}

Our $\theta$ bins span the range $0.00014-0.3$ radians.  Since the 
first zero of the kernel occurs at $k\theta = 0.013$, we expect 
sensitivity to $k$ values between $0.04 - 100 h$ Mpc$^{-1}$.  However, 
since small $k$ are the most interesting for cosmology as they probe 
the linear regime, we attempt to extend our analysis down to $k = 0.004$.
We use $18$ bins in wavenumber, $4$ per decade, spanning the range 
$0.003 - 100 h$ Mpc$^{-1}$.  Figure \ref{fig:modes} shows the recovered
power spectrum as a function of the number of modes we retain from the SV
decomposition.  We find that including $16$ SV modes gives us sensitivity 
to small $k$ but doesn't result in wild oscillations of the power spectrum.
This produces an acceptable fit to $\acf$ with $\chi^2=4.1$ with
$N_\theta - N_{SV} = 9$ degrees of freedom.

\begin{figure} [t] 
\centerline{\epsfig{file=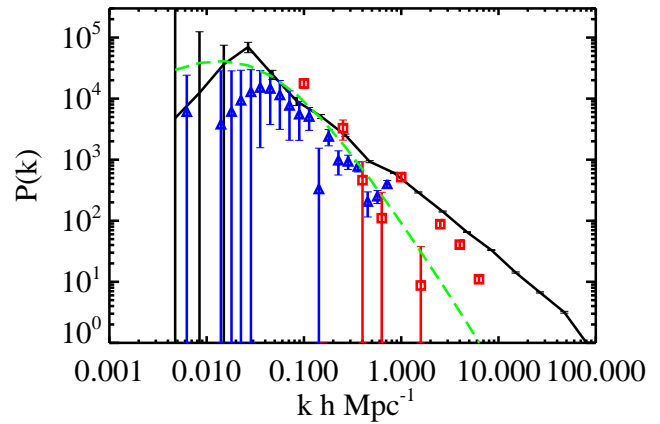, width=\linewidth}}
\caption{The power spectrum measured from 2MASS (solid line). For comparison
the results from the APM survey (triangles) as analyzed by \citet{ez:01},
and for the $r=18-19$ magnitude bin of the SDSS survey (squares) are shown.
In all cases error bars are the square root of the diagonal elements of the 
inverse covariance matrix, $\sqrt{C_{P,ii}^{-1}}$. The dashed line is the 
best fit CDM type power spectrum to the linear regime $k < 0.15$. The dotted
line is the power spectrum from the CMB as determined by \citep{sper:03}.
}\label{fig:pk}
\end{figure}

\begin{figure} [t] 
\centerline{\epsfig{file=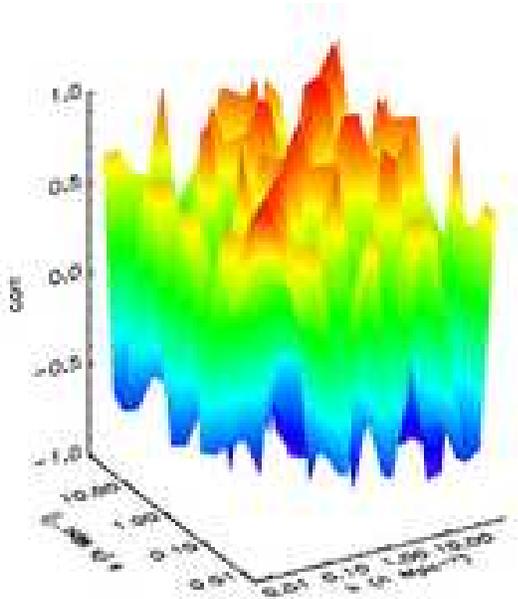, width=\linewidth}}
\caption{The correlation matrix of the power spectrum measured from 2MASS.
The correlation matrix is the covariance matrix normalized
by its diagonal elements.  One sees how different $k$ bins are not at all
independent but are all interwoven in an oscillatory fashion.  The 
elements of the correlation matrix are also given in Table \ref{tab:cov}.
}\label{fig:pkcorr}
\end{figure}

In Figure \ref{fig:pk} we plot the inferred power spectrum (with 16 SV modes)
compared to the power spectra measured for SDSS 
\citep[][$r$-band selected]{dode:02} and APM 
\citep[][$b_J$-band selected]{ez:01}. There is general agreement for the 
power spectra measured in all three surveys. The drop in power at 
$k<0.03 h$ Mpc$^{-1}$ is not statistically significant as demonstrated below. 
The plotted error bars are the square root of the diagonal elements of the 
inverse covariance matrix, $\sqrt{C_{P,ii}^{-1}}$.  Since the $k$ bins are not
independent,  they represent the uncertainty in a bin 
{\it if the values of all other bins are kept fixed}.
They are shown only for comparison to the other surveys and to get a 
feel for the relative error in a given bin.  To treat properly the uncertainty
in the measured $P(k)$ one must use the full covariance matrix.  In Table 
\ref{tab:cov} we give the inverse correlation matrix, which can be used to 
recover the inverse covariance matrix $r^{-1}$ 
(contact the authors for a more precise table).  The elements of the inverse
correlation matrix are given by
\beq
r_{ij}^{-1} = {{C^{-1}_{P,ij}}\over{\sqrt{C^{-1}_{P,ii}C^{-1}_{P,jj}}}}.
\eeq
The correlation matrix is shown in Figure \ref{fig:pkcorr}. The importance
of off diagonal elements and the oscillatory nature of different $k$ bins is
clearly seen.

\subsection{CDM Models}
In the CDM paradigm for the evolution of the Universe, the power
spectrum can be predicted based on the value of the cosmological 
parameters 
of which the most important are the matter density of the universe $\Omega_M$ 
and the Hubble constant 100$h$ Mpc$^{-1}$ kms$^{-1}$. The shape of the 
power spectrum is most sensitive to the combination $\Omega_M h$.
Along with a normalization, in terms of $\sigma_8$ for
example, this specifies a power spectrum given an initial spectral 
index $n_s$.  Thus $\Omega_M h$ and $\sigma_8$ serve as a convenient
parameterizations of the power spectrum on linear scales.  We fit $P(k)$ 
inverted from $\acf$ only for bins with $k < 0.15 h$ Mpc$^{-1}$, bins that are
still in the linear regime \citep[see][Figure 4]{perc:01}.
We use the transfer function fitting formula of \citet{eh:99}, $n_s=1.0$
and take the the baryon density $\Omega_b = 0.02h^{-2}$ in agreement 
with results from BBN \citep{kirk:03} and the CMB \citep{sper:03}
We plot the best-fit power spectrum in Figure \ref{fig:pk}
including the error in the median redshift of the sample. 
The CDM power spectrum is a good fit to the 
data giving $\chi^2/{\rm d.o.f.}=0.6$.  Hence, the apparent drop in
power on large scales, as seen by some authors \citep{gb:98,abp:01}, is not 
statistically significant. 

\begin{figure}[t] 
\centerline{\epsfig{file=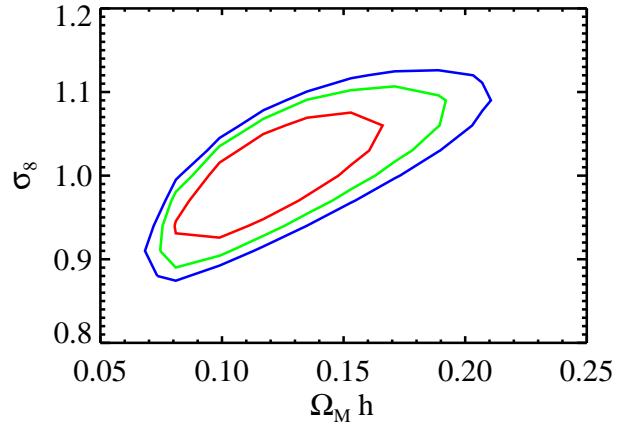, width=\linewidth}}
\caption{The $66.7\%, 95\%$ and $99\%$ confidence intervals in the 
$\sigma_8$ --- $\Omega_M h$ plane from fitting to our derived $P(k)$.
Projecting onto one-dimension gives constraints of $\sigma_8=1.0\pm0.1$
and $\Omega_M h = 0.13 \pm 0.07$.
}\label{fig:cdm}
\end{figure}

The constraints on $\sigma_8$ and $\Omega_M h$ 
are shown in Figure \ref{fig:cdm}.  We find 
$\sigma_8=1.0\pm0.1$ and $\Omega_M h =0.13 \pm 0.07$ 
($95\%$ confidence limits).  One must remember that this is only a 
parameterized fit to the power spectrum of $K_s < 13.57$ galaxies; any 
relation to cosmology requires an understanding of how galaxies trace the 
underlying dark matter density field.  The 2dfGRS find a power spectrum best
fit by $\Omega_m h = 0.2\pm0.03$ \citep{perc:01} consistent with our
measurement.

In comparison, cosmological parameters measured from the 
WMAP satellite alone give $\Omega_M h =0.19\pm0.03$, 
$\sigma_8=0.9\pm0.1$ and $n_s=.99$ \citep{sper:03}.
This would imply a $K_s$-band linear bias of 
$b_K=1.1\pm 0.2$ in good agreement with the $K_s$-band bias determined
from the 2MASS clustering dipole of $b_K=1.37\pm 0.34$ \citep{mmkw:03}. 

\section{Conclusions}
\label{sec:conc}
We have presented a measurement of the angular correlation function and 
three-dimensional power spectrum for galaxies from the 2MASS catalog.  
We have minimized the contribution of the possible contaminants of
stars, dust, seeing and sky brightness by studying their cross-correlation 
with the galaxy density and making cuts in the data until these 
cross-correlations are less then the auto-correlation signal.  These
restrictions on the data limit us to $|b| > 20\arcdeg$ and dust extinctions 
$\Delta K_S < 0.05$. More than a half million galaxies remain
for estimating the angular correlation function.

The best fit power law to 
the measured angular correlation function has a slope of
$\gam$ and an amplitude at one degree of $\A$ out to $2.5\arcdeg$.  
There are oscillations around this power law that are
statistically significant.  
The largest oscillation occurs at about 0.8 degrees,
corresponding to $700h^{-1}$ kpc at the median redshift of our survey,
as expected in halo occupation distribution descriptions of galaxy clustering
\citep{selj:00,bw:02,zehavi:04}.
The slope of this power law is in good 
agreement with other determinations of $\acf$ \citep{mes:96,conn:02}. 
We divide the sample into three magnitude 
bins and estimate the angular correlation function for each magnitude bin.
We confirm that differences in the $\acf$ amplitude at these three 
brightnesses is fully explained by differences in the median redshift, 
consistent with true clustering.  We partition the data by 
northern and southern Galactic latitude and see that at large angles
the northern Galactic latitudes show a greater correlation amplitude, 
caused by the observed large structures of the Virgo super 
cluster and the Shapley concentration in the northern Galactic hemisphere
\citep[see][Fig. 2]{mmkw:03}.
We also partition the data by effective surface brightness and find that 
galaxies with higher surface brightness are more clustered, which is a 
manifestation of the morphology-density relation \citep{dres:80}.  Their 
$\acf$ may also have a slightly steeper slope and might be evidence for
nonlocal biasing.

We solve for the three-dimensional power spectrum from the angular correlation
function using Singular Value Decomposition. Our resulting power spectrum 
is in good agreement with other measurements using the same method 
\citep{ez:01,dode:02}.  The best fit CDM power spectra gives values of 
$\Omega_M h = 0.13\pm0.07$ and $\sigma_8=1.0\pm0.1$, assuming a spectral 
index of $1.0$.  This fit is to the galaxy power spectrum and
its relation to cosmological parameters will depend on how galaxies trace
the underlying mass distribution.  The ratio of our measured $\sigma_8$ to
that determined from WMAP \citep{sper:03} would imply a $K_s$ band linear 
bias of $b_K=1.1\pm0.2$.

Our $K_s$-band selected power spectrum, is in good agreement with 
other determinations of the power spectrum in surveys selected in other 
wavebands in the linear regime. On smaller length scales there is a 
significant difference between other galaxy power spectra and that measured
for $K_s$-band selected galaxies.  Thus the power spectrum
measured here, in combination with the power spectrum measured from the CMB
and other large galaxy surveys, will enable one to place strong constraints on
theories of galaxy formation.

\acknowledgments
We are grateful for helpful discussions and correspondence about the 2MASS
catalog with Roc Cutri, Tom Jarrett, Steve Schneider, and Rae Stiening. We 
thank Rennan Barkana, James Bullock, John Peacock, Risa Wechsler and 
David Weinberg for 
conversations about the three dimensional power spectrum.  We thank the
anonymous referee for helpful comments that improved the paper.
We acknowledge support by JPL/NASA through the 2MASS core science projects,
by NASA grants NAG5-12038 and NAG5-13102, and by NSF grants AST-9988146 and
AST-0205969.
This publication makes use of data products 
from the Two Micron All Sky Survey, which is a joint project of the 
University of Massachusetts and the Infrared Processing and Analysis 
Center/California Institute of Technology, funded by the National 
Aeronautics and Space Administration and the National Science Foundation.

This publication also makes use of the {\it Sloan Digital
Sky Survey} (SDSS).
Funding for the creation and distribution of the SDSS
Archive has been provided by the Alfred P.\ Sloan Foundation, the
Participating Institutions, the National Aeronautics and Space Administration,
the National Science Foundation, the US Department of Energy,
the Japanese Monbukagakusho, and the Max Planck Society.  The SDSS
Web site is {\texttt http://www.sdss.org/}.  The SDSS Participating
Institutions are the University of Chicago, Fermilab, the Institute
for Advanced Study, the Japan Participation Group, the Johns Hopkins
University, the Max Planck Institut f\"ur Astronomie, the Max
Planck Institut f\"ur Astrophysik, New Mexico State University,
Princeton University, the United States Naval Observatory, and
the University of Washington.
This publication also made use of NASA's Astrophysics Data System
Bibliographic Services.

\bibliographystyle{apj}

\bibliography{cor,2mass,cosmo,me}

\pagestyle{empty}
\begin{turnpage}
\begin{table*}
\begin{center} 
\begin{tabular}{ccccccccccccccccccc}
$k$ & 0.0047 & 0.0084 &  0.015 &  0.027 &  0.047 &  0.084 &
   0.15 &   0.27 &   0.47 &   0.84 &    1.5 &    2.7 &
    4.7 &    8.4 &    15. &    27. &    47. &    84.\\
$P(k)$ &  4792.2 & 12419.1 & 36534.2 & 69942.7 & 24786.4 &  9083.5 &
  5107.2 &  2508.1 &   928.1 &   606.5 &   290.9 &   141.6 &
    65.2 &    33.1 &    14.2 &     6.7 &     3.2 &     0.9\\
$\sqrt{C_{P,ii}^{-1}}$ & 2.9E-06 & 8.9E-06 & 2.6E-05 & 7.4E-05 & 2.3E-04 & 7.2E-04 &
 2.6E-03 & 8.4E-03 & 2.7E-02 & 6.6E-02 & 1.5E-01 & 3.5E-01 &
 1.1E+00 & 1.9E+00 & 2.8E+00 & 6.1E+00 & 9.3E+00 & 8.1E+00\\
 &  1.00 &  1.00 &  0.98 &  0.83 &  0.51 &  0.43 &
  0.01 & -0.32 & -0.05 & -0.18 & -0.25 &  0.16 &
  0.43 &  0.08 & -0.16 & -0.41 & -0.29 &  0.26\\
 &  1.00 &  1.00 &  0.99 &  0.86 &  0.55 &  0.44 &
  0.02 & -0.31 & -0.07 & -0.21 & -0.25 &  0.18 &
  0.43 &  0.06 & -0.17 & -0.39 & -0.27 &  0.25\\
 &  0.98 &  0.99 &  1.00 &  0.92 &  0.65 &  0.48 &
  0.06 & -0.30 & -0.13 & -0.29 & -0.26 &  0.23 &
  0.42 & -0.02 & -0.19 & -0.34 & -0.20 &  0.21\\
 &  0.83 &  0.86 &  0.92 &  1.00 &  0.87 &  0.58 &
  0.16 & -0.25 & -0.28 & -0.50 & -0.27 &  0.33 &
  0.36 & -0.20 & -0.20 & -0.18 & -0.02 &  0.11\\
 &  0.51 &  0.55 &  0.65 &  0.87 &  1.00 &  0.77 &
  0.30 & -0.18 & -0.38 & -0.59 & -0.35 &  0.28 &
  0.36 & -0.27 & -0.14 & -0.08 &  0.04 &  0.10\\
 &  0.43 &  0.44 &  0.48 &  0.58 &  0.77 &  1.00 &
  0.57 & -0.14 & -0.22 & -0.29 & -0.49 &  0.08 &
  0.59 &  0.07 & -0.02 & -0.33 & -0.32 &  0.33\\
 &  0.01 &  0.02 &  0.06 &  0.16 &  0.30 &  0.57 &
  1.00 &  0.36 & -0.06 & -0.12 & -0.23 &  0.04 &
  0.16 &  0.09 &  0.05 & -0.00 & -0.01 &  0.02\\
 & -0.32 & -0.31 & -0.30 & -0.25 & -0.18 & -0.14 &
  0.36 &  1.00 &  0.41 & -0.14 & -0.11 & -0.18 &
 -0.29 & -0.01 &  0.57 &  0.45 &  0.02 & -0.50\\
 & -0.05 & -0.07 & -0.13 & -0.28 & -0.38 & -0.22 &
 -0.06 &  0.41 &  1.00 &  0.59 & -0.11 & -0.47 &
 -0.03 &  0.44 &  0.29 &  0.01 & -0.29 & -0.01\\
 & -0.18 & -0.21 & -0.29 & -0.50 & -0.59 & -0.29 &
 -0.12 & -0.14 &  0.59 &  1.00 &  0.04 & -0.42 &
  0.06 &  0.62 &  0.02 & -0.26 & -0.29 &  0.20\\
 & -0.25 & -0.25 & -0.26 & -0.27 & -0.35 & -0.49 &
 -0.23 & -0.11 & -0.11 &  0.04 &  1.00 &  0.31 &
 -0.37 & -0.16 &  0.06 &  0.24 &  0.11 & -0.14\\
 &  0.16 &  0.18 &  0.23 &  0.33 &  0.28 &  0.08 &
  0.04 & -0.18 & -0.47 & -0.42 &  0.31 &  1.00 &
  0.33 & -0.23 & -0.06 & -0.02 &  0.03 & -0.16\\
 &  0.43 &  0.43 &  0.42 &  0.36 &  0.36 &  0.59 &
  0.16 & -0.29 & -0.03 &  0.06 & -0.37 &  0.33 &
  1.00 &  0.48 & -0.02 & -0.57 & -0.53 &  0.28\\
 &  0.08 &  0.06 & -0.02 & -0.20 & -0.27 &  0.07 &
  0.09 & -0.01 &  0.44 &  0.62 & -0.16 & -0.23 &
  0.48 &  1.00 &  0.22 & -0.52 & -0.52 &  0.14\\
 & -0.16 & -0.17 & -0.19 & -0.20 & -0.14 & -0.02 &
  0.05 &  0.57 &  0.29 &  0.02 &  0.06 & -0.06 &
 -0.02 &  0.22 &  1.00 &  0.27 & -0.34 & -0.44\\
 & -0.41 & -0.39 & -0.34 & -0.18 & -0.08 & -0.33 &
 -0.00 &  0.45 &  0.01 & -0.26 &  0.24 & -0.02 &
 -0.57 & -0.52 &  0.27 &  1.00 &  0.49 & -0.76\\
 & -0.29 & -0.27 & -0.20 & -0.02 &  0.04 & -0.32 &
 -0.01 &  0.02 & -0.29 & -0.29 &  0.11 &  0.03 &
 -0.53 & -0.52 & -0.34 &  0.49 &  1.00 & -0.35\\
 &  0.26 &  0.25 &  0.21 &  0.11 &  0.10 &  0.33 &
  0.02 & -0.50 & -0.01 &  0.20 & -0.14 & -0.16 &
  0.28 &  0.14 & -0.44 & -0.76 & -0.35 &  1.00\\

\hline
\end{tabular}
\caption{The results of the best fit power spectrum using 16 SV modes. 
The first row are the centers of the $k$ bins in $h$ Mpc$^{-1}$.  The second
row are the best fit values of $P(k)$.  The third row is the square root
of the covariance matrix, $\sqrt{C_{P,ii}}$.  The fourth row is the square
root of the inverse covariance matrix $\sqrt{C_{P,ii}^{-1}}$ and the rest of 
the table gives the $18 \times 18$ symmetric inverse correlation matrix 
$r^{-1}$ (see text).  Note redshift errors are not included in the above.
For a table with greater precision please contact the authors.
}\label{tab:cov}
\end{center}
\end{table*}  
\end{turnpage}

\end{document}